\documentclass[
onecolumn,
 amsmath,amssymb,
 aps,
]{revtex4-2}

\usepackage{graphicx}
\usepackage{dcolumn}
\usepackage{bm}
\usepackage{siunitx}
\usepackage{booktabs}
\usepackage{xcolor}

\usepackage[
margin=1.4in,
]{geometry}

\DeclareMathOperator{\tr}{tr}
\newcommand{\dd}{\mathrm{d}}
\newcommand{\bff}{\boldsymbol{f}}
\newcommand{\br}{\boldsymbol{r}}
\newcommand{\bu}{\boldsymbol{u}}
\newcommand{\bx}{\boldsymbol{x}}
\newcommand{\btheta}{\boldsymbol{\theta}}
\newcommand{\bsigma}{\boldsymbol{\sigma}}
\newcommand{\btau}{\boldsymbol{\tau}}
\newcommand{\srate}{\dot{\gamma}}
\newcommand{\bA}{\boldsymbol{\mathrm{A}}}
\newcommand{\bD}{\boldsymbol{\mathrm{D}}}
\newcommand{\bI}{\boldsymbol{\mathrm{I}}}
\newcommand{\bO}{\boldsymbol{\mathrm{O}}}
\newcommand{\RR}{\mathbb{R}}

\begin{document}


\title{Non-Newtonian viscous fluid models with learned rheology accurately reproduce  \\ Lagrangian sea ice simulations}

\author{Gonzalo G. de Diego}
\email{gg2924@nyu.edu}
\author{Georg Stadler}
\affiliation{Courant Institute of Mathematical Sciences, New York University, 251 Mercer Street, New York City, 10012, NY, USA}


\begin{abstract}
Polar sea ice is crucial to Earth's climate system. Its dynamics also affect coastal communities, wildlife, and global shipping. 
Sea ice is typically modeled as a continuum fluid using a model proposed almost 50 years ago, which is moderately accurate for packed ice, but loses its predictive accuracy outside of the central ice pack.
Discrete element methods (DEMs), which are commonly used for modeling granular media, offer an alternative by resolving the behavior of individual ice floes, including collisions, frictional contact, fracture, and ridging. However, DEMs are generally too costly for large-scale simulations.
To address this, we present a framework for inferring rheological behavior from DEM velocity data. We characterize isotropic constitutive laws as scalar functions of the principal invariants of the strain-rate tensor. These functions are parameterized by neural networks trained on DEM data. By combining machine learning and finite element methods, we incorporate the governing partial differential equation (PDE) into the training, requiring to solve a PDE-constrained optimization problem for the network parameters.
We focus on unidirectional parallel shear flows, which allow us to infer the effective shear viscosity. We find that, over a wide range of ice concentrations, the velocity fields observed in a complex sea ice DEM can be captured by a nonlinear rheology. Depending on the ice concentration, a shear-thinning or a shear-thickening behavior is observed. Moreover, the effective shear viscosity is found to increase by several orders of magnitude with changes as small as 5\% in the sea ice concentration.
We show that the learned rheology generalizes to different forcing scenarios, time-dependent problems, and settings in which compressibility is not a dominant factor. For these reasons, our framework represents a major step towards developing non-Newtonian models that accurately reproduce observed sea ice dynamics. 
\end{abstract}
\maketitle


\section{Introduction}
Sea ice plays a vital role in Earth's climate. Covering about 10\% of the ocean's surface at its maximum extent \cite{feltham2008, weeks2010}, its high albedo plays an important role in Earth's energy budget \cite{kwok2011}. Key oceanographic processes are also driven by sea ice. For example, dense water masses produced by sea ice formation off the coasts of Antarctica contribute to the formation of the Antarctic Bottom Water, a crucial water mass in the Southern Ocean that transports heat and carbon throughout Earth's oceans \cite{zhou2023}. 

Accurate predictions of the dynamics of the sea ice cover remain an important challenge in Earth System Models \cite{foxkemper2023}. Mathematical models for sea ice fall into two broad categories: Lagrangian approaches that track individual ice floes, and continuum models for sea ice based on partial differential equations (PDEs). Lagrangian approaches use conservation of momentum and angular momentum, together with floe-level processes, such as frictional collisions, and parameterizations of fracturing and ridging, to describe the evolution of each ice floe. Numerical methods that resolve the motion of a large number of particles are known as discrete element methods (DEMs). While considered a promising avenue for sea ice modeling, the use of DEMs in large-scale simulations is prohibitive due to high computational costs \citep{blockley2020}.

The challenge of DEMs for large-scale sea ice modeling can be addressed by using continuum models of the ice cover. Continuum models, which have been the most common approach for modeling sea ice on large scales ($\sim 100\,\mathrm{km}$ and larger) \cite{blockley2020}, are based on PDEs. Their computational realization is, in general, much less demanding than that of DEMs. However, the use of phenomenological parameterizations to represent complex physical phenomena that can rarely be verified with observations limits the success of continuum models \cite{feltham2008}. The relationship between stress and deformation is the most crucial material law in a continuum model for sea ice \cite{rothrock1975, feltham2008, blockley2020}. This relationship, known as rheology in the context of viscous fluids, parameterizes the internal stress field that emerges from mechanical interactions between ice floes. The most widely used rheological model in sea ice is the Hibler model \cite{hibler1979}, which is a viscous-plastic model based on the plastic model developed by the AIDJEX group \cite{coon1974, rothrock1975} 50 years ago. Here, internal stresses are assumed to be the result of pressure ridges formed between ice floes under compression. Despite its success in reproducing certain observational features \cite{blockley2020, roach2020}, Hibler's model (and its variations used in Earth System Models) has severe limitations, in particular outside the central ice pack, such as in the marginal ice zone \cite{dumont2022}. In the marginal ice zone, where the sea ice concentration is lower than in the central ice pack, sea ice dynamics is dominated by granular effects (collisions and enduring frictional contact) that are not included in Hibler's model \cite{herman2022}.  

Data-driven approaches based on scientific machine learning (sciML) have emerged as a promising avenue for parameterizing physical processes \cite{lei2020, lu2021, shamekh2023, kochkov2024, wang2025}.
For viscous fluids (generally non-Newtonian), data-driven approaches circumvent the need to derive a rheology phenomenologically or from first principles, instead using data to learn the rheology. A large class of methods assume a functional form for the rheology which includes unknown parameters. These unknown parameters are then fitted using simulation or observational data \cite{mahmoud2021, hu2024}. However, using a 
functional form for the rheology requires prior knowledge of physics and, even if available, can limit the generality of these methods. In recent years, a small number of works have employed ideas from sciML to overcome this limitation. For example, sciML techniques have been used to construct rheological laws in terms of a given library of functions \cite{saadat2022, mahmoud2024, thakur2024}. Another class of sciML methods, particularly relevant to this work, represents a rheological model directly as a neural network (NN) \cite{reyes2021, lennon2023, parolini2025}. Representing a fluid's rheology with an NN enables a form-agnostic approximation that can learn complex rheologies whenever sufficiently rich training-datasets are available. Moreover, restricting the use of NNs to unknown terms in the fluid's rheology enables a high degree of interpretability because the NNs represent objective relationships between physical quantities \cite{lennon2023}. 
This is in contrast to e.g.~representing the solution operator to a PDE with a large NN \cite{kovachki2023}.

In this work, we infer a
continuum non-Newtonian viscous fluid model that reproduces the velocity
fields computed with a DEM for sea ice called {\it SubZero}
\cite{manucharyan2022}. This DEM evolves irregularly polygonal-shaped
ice floes that interact through collisions, friction, ridging, and
fracture, allowing it to successfully capture observed statistical properties, such as the power-law appearance of the floe size distribution and the long-tailed ice thickness distribution. To our knowledge, no systematic inference of the rheology
for such an intricate DEM ice model has been attempted.
\emph{SubZero}'s stress-strain data gives no clear indication of the
existence of an underlying rheology. This is in contrast to existing
work that also uses ML-based rheology parameterizations to target
fluids whose dynamics are expected to be captured with standard,
although often complex, nonlinear viscosity models \cite{mahmoud2021,
  reyes2021, saadat2022, lennon2023, mahmoud2024, parolini2025}. In
fact, the potential lack of a rheological model fitting the data
motivates the development of a novel training strategy. Unlike most
approaches found in the literature, which only consider stress-strain
data \cite{lennon2023, mahmoud2024, parolini2025}, we train our NN by
minimizing the misfit between the DEM's velocity data and the
continuum velocity field. This requires solving the continuum model
with an NN-based parameterization of the effective viscosity, and
combining adjoint-based PDE and backpropagation sciML techniques to
compute gradients of our misfit. Similarly to \cite{lennon2023,
  parolini2025}, we tailor our NN
to ensure that the continuum model satisfies key physical and
mathematical principles. For example, we guarantee frame-indifference
by characterizing our rheology in terms of the principal invariants of
the strain-rate tensor, and enforce monotonicity of a certain function
to ensure the continuum model is uniquely solvable.
This results in a physically-sensible
continuum model that reproduces the DEM at a much reduced
computational cost.

\begin{figure*}[tb]
    \centering
    \includegraphics[width=\textwidth]{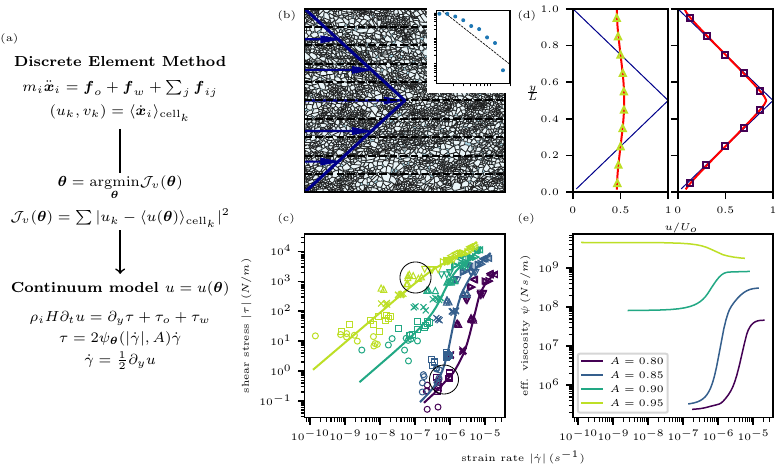}
    \caption{(a) Diagram summarizing the training framework: with the DEM we compute the position $\bx_i$ for floe $i$ with mass $m_i$ forced with ocean drag ($\bff_o$), wind drag ($\bff_w$), and collisions ($\bff_{ij}$). We extract horizontal velocities $u_k$ averaged over each cell $k$ and minimize the mismatch $\mathcal{J}_v$ with the solution $u(\btheta)$ to the continuum model, which is defined in terms of an NN-based rheology $\psi_{\btheta}$ that also depends on the sea ice concentration $A$. (b) Ice-floe field simulated in \emph{SubZero}. For generating the training data, we drive the floes with the hat-shaped horizontal ocean velocity profile shown in blue and zero wind velocity. We average over the elongated cells with black dashed edges. Inset: realistic power-law floe size distribution satisfied by ice floes for areas between $10^{-4}L^2$ and $2\times 10^{-3}L^2$. (c) NN-based shear stress model inferred from training are shown using solid lines. Markers represent DEM data. (d) Two steady velocity profiles from training dataset. Sea ice NN model in red line and DEM data in markers (markers of same type and color in panel (c), circled, correspond with same steady states); blue lines represent ocean velocity. (e) Effective shear viscosity $\psi_{\btheta}$ inferred with training.}
    \label{fig:setup}
\end{figure*}

\section{A framework for rheology inference}\label{sec:framework}

The determination of a rheology for sea ice is a long-standing
challenge in climate modeling \cite{feltham2008, blockley2020}. Here, we introduce a
data-driven approach for discovering a concentration-dependent
rheological law for sea ice. We apply this method to DEM simulation data generated
with \emph{SubZero} in a unidirectional parallel flow 
configuration where the sea ice variables only vary along the vertical dimension. This setup allows us to infer an effective shear viscosity
relating the shear stress to the shear strain-rate. By
representing the rheological model with a neural network (NN) and
embedding key physical principles into this parameterization, our approach avoids having to choose a
specific functional form for the rheology. This flexibility
is indispensable for modeling a material
of such complexity as sea ice. Figure~\ref{fig:setup} provides an
overview of our framework, including the equations underlying data
generation and rheology learning (a), and the inferred
rheologies for different ice concentrations (e), which can be
studied and interpreted using classical non-Newtonian fluid dynamic theory.

To define what form physically meaningful material relations can take,
the principle of isotropic frame-indifference is a fundamental constraint on a
rheological law \cite{bampi1980}. Essentially, this principle states that the
constitutive laws describing a material's behavior should be
independent of the frame of reference.
The set of models that satisfy
this principle can be fully characterized in terms of tensor invariants, as explained in Appendix \ref{appsubsec:isotropic_rheology}.
In one dimension, a general expression for a frame-indifferent rheology for
sea ice that also depends on the sea ice concentration $A$ is given by
\begin{align}\label{eq:rheology}
    \tau = 2 \psi(|\srate|, A) \srate.
\end{align} 
Here, $\tau$ is the shear stress, $\srate$ the shear strain-rate, and $\psi$ is the effective shear viscosity. The shear stress $\tau$ and strain-rate $\srate$ correspond with the off-diagonal terms in the Cauchy stress and the strain-rate tensors, respectively, as clarified in appendix \ref{appsec:continuum_model}. The internal stresses in sea ice also depend on e.g.~the ice thickness $H$ and floe size; for simplicity, we ignore the dependence of the effective viscosity on additional variables and assume the sea ice thickness to remain uniform in the DEM and the continuum model. To remain form-agnostic, we represent the effective shear viscosity in terms of feedforward NNs, that is, 
the function $\psi=\psi_{\btheta}$ is parameterized by network weights $\btheta$. 
We experimented with different parameterizations for $\psi$, including polynomials, and found NNs to work best when enforcing the monotonicity condition described in appendices \ref{appsubsec:well_posedness} and \ref{appsec:optim}.

The weights $\btheta$ are inferred from data that we generate with 
a DEM for sea ice that represents ice floes with realistic polygonal shapes \cite{manucharyan2022}.
For simplicity, we assume that floes do not deform mechanically (i.e.~floes are not allowed to ridge or fracture) in the DEM simulations. 
Including them in future work is straightforward. Computations in the DEM are initialized by generating a floe field with a packing algorithm based on a Voronoi tesselation of the domain. As detailed in Appendix \ref{appsec:DEM}, we have implemented a custom-made tesselation algorithm that results in a power-law floe-size distribution with a slope of $m\approx -1.75$, following reports from observed satellite imagery \cite{denton2022}, as depicted in panel (b) of figure~\ref{fig:setup}. The ice floes are then evolved in time by solving equations for the conservation of momentum and angular momentum for each floe.

Two sets of data can be computed from DEM simulations. At each time step, we extract horizontal velocities $u_k$ and shear stresses $\tau_k$ by spatially averaging these quantities over a grid with cells indexed by $k$. In this way, we map the DEM's floe-based Lagrangian data into a Eulerian representation of sea ice variables that is compatible with a continuum model. In the DEM, the Cauchy stress tensor for a floe with index $i$ is computed in terms of collisional forces using the (symmetric) Love-Weber formula:
\begin{align}\label{eq:love-weber}
    \bsigma_{i} := \frac{1}{2a_i}\sum_j \left( \bff_{ij}\otimes\br_{ij} + \br_{ij}\otimes\bff_{ij}\right),
\end{align}
see \cite{nicot2013} for a derivation. In \eqref{eq:love-weber}, $a_i$ represents the area of floe $i$, $\bff_{ij}$ the $j$-th contact force, and $\br_{ij}$ the vector connecting the $j$-th contact point with the floe's center of mass.

By computing the strain-rate $\srate_k$ with the averaged DEM's velocity field, we generate dynamic and kinematic datasets, given by $\mathcal{D} = \{ (\tau_k,\srate_k) \}$ and $\mathcal{K} = \{ u_k \}$, respectively. This correspondence between dynamic and kinematic quantities is expressed in figure~\ref{fig:setup} with circled stress-strain points in panel (c) corresponding to the averaged horizontal velocity points in panel (d). These two data sets suggest that two different misfit functionals can be defined to learn the rheology by training our NNs. One can either minimize the stress misfit given by
\begin{align}\label{eq:stress-misfit}
    \mathcal{J}_s(\btheta) := \sum_{(\tau_k,\srate_k)\in\mathcal{D}} |\log(|\tau_k|) -  \log(|2 \psi_{\btheta}(|\srate_k|, A) \srate_k|)|^2,
\end{align}
where we use a logarithmic mean squared residual to account for large changes in stress. Alternatively, for a continuum model that produces a horizontal velocity field $u$ given NN weights $\btheta$ and a sea ice concentration $A$, we can define the velocity misfit 
\begin{align}\label{eq:vel-misfit}
    \mathcal{J}_v(\btheta) := \sum_{u_k\in\mathcal{K}} |u_k -  \langle u(\btheta, A)\rangle_k|^2.
\end{align}
Here, $u = u(\btheta, A)$ is the solution to the continuum model for given NN parameters $\btheta$ and sea ice concentration $A$, and $\langle \cdot\rangle_k$ denotes averaging over cell $k$. The continuum model 
is given by the rheological law \eqref{eq:rheology} and an equation for conservation of momentum, which in one dimension is
\begin{align}\label{eq:conservation_of_momentum}
    \rho_i H \partial_t u - \partial_y \tau = \tau_o(u) + \tau_w.
\end{align}
%
Here, $\rho_i$ is the density of ice, $\tau_o$ and $\tau_w$ are the drag forces due to the ocean and the wind, respectively, and $\tau$ is the shear stress. The drag forces are defined in terms of the ocean and wind velocities $u_o$ and $u_w$, the densities $\rho_o$ and $\rho_a$, and the drag coefficients $C_o$ and $C_a$, respectively, as follows: 
\begin{align}\label{eq:defn_tau}
    \tau_o(u) := \rho_o C_o |u_o - u|(u_o-u) \quad \text{and} \quad \tau_w(u) := \rho_a C_a |u_w - u|(u_w-u).
\end{align}
Since $|u_w| \gg |u|$ in general, one uses 
\begin{align}\label{eq:defn_tauw}
    \tau_w := \rho_a C_a |u_w|u_w,
\end{align}
a common practice which we take up in this work.
In Appendix \ref{appsec:continuum_model}, we derive \eqref{eq:conservation_of_momentum} from a general two-dimensional equation for sea ice. There, we see that $\tau_o(u)$ and $\tau_w$ represent the horizontal components of the two-dimensional vector fields $\btau_o$ and $\btau_w$, respectively, in the one-dimensional configuration introduced in section \ref{appsubsec:deriving_one_dim_model}. Since we assume $A$ and $H$ to be spatially constant, \eqref{eq:rheology} and \eqref{eq:conservation_of_momentum}, together with $\srate = \frac{1}{2}\partial_yu$, constitute the closed system of equations implicitly solved in the solution map from $(\btheta,A)$ to $u = u(\btheta,A)$. This PDE contains an NN-parametrized function $\psi_{\btheta}$. To solve it numerically, we use the finite element module Firedrake \cite{ham2023}. Firedrake's pyadjoint module \cite{mitusch2019} utilizes automatic differentiation capabilities to efficiently compute derivatives of a functional such as $\mathcal{J}_v$ using adjoint variables. Moreover, recent developments have achieved an effective coupling between Firedrake and PyTorch that allows us to work with the NN-based operator $\psi_{\btheta}$ \cite{bouziani2024}. Our implementation based on Firedrake and PyTorch can be found in the following repository \cite{RheoDis}.   

As depicted in figure~\ref{fig:setup}, in this work we compute the NN-weights $\btheta$ by minimizing $\mathcal{J}_v$, i.e., we
infer a rheology for sea ice only from velocity data. This approach has several advantages. Firstly, unlike stress-strain data, velocity data for sea ice is readily available from satellite imagery, enabling the use of real data to infer a rheological model. Secondly, the stress-strain data generated from the DEM through cell averaging is noisy, see panel (c) in figure~\ref{fig:setup}. For this reason, it is unclear whether there is an underlying rheology. In fact, in the next section we find that a model that closely fits the stress-strain data does not approximate the DEM's velocity fields accurately in all regimes. From a computational perspective, the efficient minimization of $\mathcal{J}_v$ requires the computation of gradients of the PDE-based map $(\btheta,A)\mapsto u(\btheta, A)$. For this, we need to solve an NN-based PDE and its corresponding linearized adjoint problem numerically. In contrast, the minimization of $\mathcal{J}_s$ corresponds to a standard non-linear regression problem.

A major challenge that arises when minimizing $\mathcal{J}_v$ is the need for further restrictions on the rheological model $\psi_{\btheta}$ that guarantee the existence of unique solutions to the PDE $u = u(\btheta, A)$. For a general function $\psi_{\btheta}$, we cannot expect solutions to our continuum model to be unique or even exist. Under these conditions, any optimization algorithm for minimizing $\mathcal{J}_v$ is severely impaired because the map $(\btheta,A)\mapsto u(\btheta, A)$ is likely ill-defined. We remedy this by enforcing two additional properties on $\psi_{\btheta}$. Firstly, we enforce $\psi_{\btheta} \geq 0$ by using an ELU activation unit increased by one in the last layer of the NN $\chi$, introduced below in \eqref{eq:NN_form}. The non-negativity of $\psi_{\btheta}$ implies that internal stresses are always dissipative. Secondly, we require the map $s\mapsto \psi_{\btheta}(|s|, A)s$ to increase monotonically. As explained in Appendix \ref{appsubsec:well_posedness}, when this condition holds, the continuum model is equivalent to the minimization of a strictly convex energy functional, which one can expect to have a unique minimizer. For this reason, we penalize negative values of the derivative of the map $s\mapsto \psi_{\btheta}(|s|, A)s$ when training our neural network; see Appendix \ref{appsubsec:penalty} for details. 

\section{Results}

The numerical results in this section describe the learning and testing for generalization of our rheology model. 
The material parameters used in the computations for the DEM and the continuum model can be found in Table \ref{tab:material_parameters}. The three movies, labeled S1 to S3 and included as supplemental material to this article, show DEM simulations representative of the training and testing problems. In movie S1, a problem from the training set presented in section \ref{subsec:training} is presented; a floe-field with concentration $A=0.9$ is driven by the hat-shaped ocean profile in figure \ref{fig:setup} with a maximum magnitude of $\SI{1}{\meter\per\second}$ and no wind. Movie S2 depicts the unsteady problem considered in section \ref{subsubsec:1d_tests}, where a randomly generated and time-dependent wind velocity drives the ice floes. Finally, movie S3 represents the two-dimensional test from section \ref{subsubsec:2d_test}.

\subsection{Training the neural network}\label{subsec:training}

\begin{figure*}[t]
    \centering
    \includegraphics[width = 0.7\textwidth]{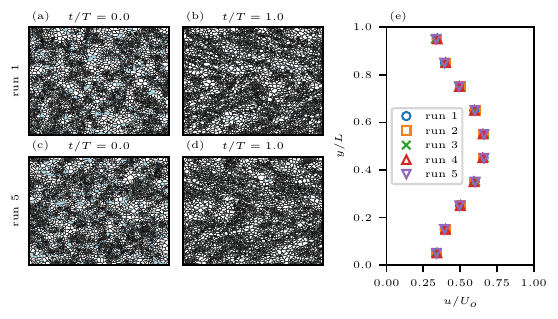}
    \caption{Steady states computed with the DEM for $A = 0.85$ and $U_o = \SI{1}{\meter\per\second}$ with five randomized floe-field initializations (runs 1-5). Initial and final floe fields for run 1 (a)-(b) and run 2 (c)-(d). (e) Velocity profiles computed with runs 1-5.}
    \label{fig:diff_init}
\end{figure*}

\begin{figure*}[t]
    \centering
    \includegraphics[width=0.9\textwidth]{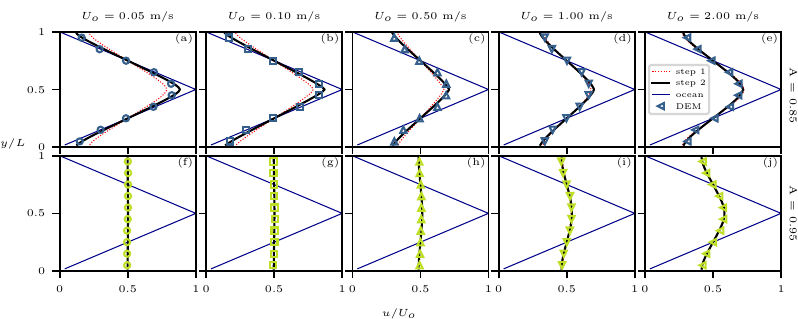}
    \caption{Comparison between velocity profiles computed with learned continuum model (black lines) and DEM (markers) for training. The steady states are computed with the DEM for concentrations $A = 0.85$ (a-e) and $A = 0.95$ (f-j) and maximum ocean velocities $U_o$ between 0.05 and 2 m/s. The rheology of the continuum models is inferred in a two step optimization process: in step 1, we minimize the stress-strain misfit $\mathcal{J}_s$ (red dotted line) and, in step 2, using the fit found in step 1 as initial guess, the velocity misfit $\mathcal{J}_v$ (black line), yielding the final continuum model.}
    \label{fig:velocity}
\end{figure*}

We represent our rheological model $\psi_{\btheta}$ in terms of two NNs, $\xi$ and $\chi$, such that
\begin{align}\label{eq:NN_form}
    \psi_{\btheta}(|\srate|, A) = e^{\xi(A)} \chi(|\srate|, A).
\end{align}
The exponential term acts as a scaling factor that accounts for large changes in the effective viscosity with the concentration. The two feedforward NNs we use contain 2 hidden layers with 5 neurons each. We train the model $\psi_{\btheta}$ by computing steady states with the DEM to the problem depicted in panel (b) of figure~\ref{fig:setup}. For the DEM computations, we consider a square domain with length $L = \SI{100}{\kilo\meter}$ and periodic boundary conditions in both directions. A horizontal ocean current with a triangular-shaped profile whose maximum value is denoted by $U_o$ drives the ice floes from east to west; no wind is included in the training setup. In all DEM simulations performed in this work, we use 5000 floes with thickness $H = \SI{2}{\meter}$, which follow a realistic floe size distribution as discussed in Appendix \ref{appsec:DEM}. We consider four different concentrations ($A = 0.8$, $0.85$, $0.9$ and $0.95$) and seven maximum ocean velocities ($U_o = 0.05$, $0.1$, $0.25$, $0.5$, $1$, $1.5$, $\SI{2}{\meter\per\second}$). We run the simulations to time $T$ such that a quasi-steady state is reached over the last 20\% of the time steps. We judge the quasi-steady to have been reached whenever the norm $\|u(t) - u(T)\|$ remains approximately constant, on average. In practice, we find this to be the case when $T = \SI{2e5}{\second}$. To ensure that the steady results we find are independent of the initial floe field configuration, we compute steady states for $A = 0.85$ and $U_o = \SI{1}{\meter\per\second}$ with five different randomized initializations. Our results, depicted in figure \ref{fig:diff_init}, indicate that the resulting velocity profiles are visibly indistinguishable.

Once a steady state has been reached, we extract dynamic and kinematic data for these steady states by averaging these quantities in time, over the last 20\% time steps, and spatially, over ten uniformly spaced cells that span the horizontal length of the domain. These cells are depicted in figure \ref{fig:setup}.A and contain, on average 500 floes each. With this data, we optimize for the network weights $\btheta$ in two steps: first, we minimize the stress misfit objective $\mathcal{J}_s$, and second, we minimize the velocity misfit $\mathcal{J}_v$. In the second step, we use the weights $\btheta$ from the first step as initialization for the optimization algorithm. A more detailed account of the optimization algorithm is provided in Appendix \ref{appsec:optim}. 

Panels (c) and (d) in figure~\ref{fig:setup} show the rheological relationship resulting from the minimization of $\mathcal{J}_v$. The discovered rheology closely follows the DEM's stress-strain data, despite the fact that these data points were not used in the second stage of the optimization. The effective viscosity increases substantially with the concentration, reflecting the increase in internal stresses as the ice pack becomes more dense. Moreover, for all concentrations but $A = 0.95$, sea ice exhibits a shear-thickening behavior. Further information on the training process can be found in figures \ref{fig:optimization} and \ref{fig:rheology} in appendix \ref{appsec:optim}. The values of the misfit functionals $\mathcal{J}_s$ and $\mathcal{J}_v$ at each iteration of the optimization are plotted in figure~\ref{fig:optimization}. In figure~\ref{fig:rheology}, we complement panel (e) of figure~\ref{fig:setup} by additionally showing the rheologies resulting from step 1 of the optimization procedure.

The velocity profiles computed with the models resulting from steps 1 and 2 of our optimization routine are shown in figure~\ref{fig:velocity} for two concentrations. For lower concentrations and slower ocean currents, we find that an accurate stress-strain fit does not correspond to an accurate sea ice velocity. Figures \ref{fig:optimization} and \ref{fig:rheology} also show that the largest changes from step 1 to step 2 of the optimization occur for $A = 0.85$; conversely, almost no changes are observed for $A = 0.95$. These results demonstrate the utility of our velocity-based optimization approach when working with DEM data. Figure~\ref{fig:velocity} also contains visual proof of the emergence of a shear-thickening and thinning rheology for low and high concentrations, respectively. For $A = 0.85$, the non-dimensionalized velocity profile $u/U_o$ flattens as $U_o$ increases. In contrast, the converse can be observed for $A = 0.95$. A flattening of the velocity profile is an indication of the material's strengthening with an increasing strain-rate.

\begin{figure*}[t]
\centering
\includegraphics[width=0.85\textwidth]{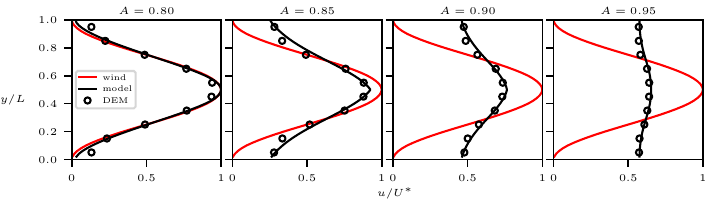}
\caption{Horizontal velocity profiles for a steady one-dimensional problem used for testing the model's generalizability. We plot the velocity computed with the DEM (markers) and with the learned model (black lines). The red line represents the wind velocity profile, which is given by a cosine profile $u_w(y) = U_w/2 (1 - \cos(2\pi y/L))$ with amplitude $U_w = \SI{20}{\meter\per\second}$.}
\label{fig:wind}
\end{figure*}

\subsection{Testing the generalizability of our rheological model}
To test our model's capabilities in capturing the DEM's velocities, we use several problems that differ substantially from our training configuration. We emphasize that the network weights $\btheta$ are fixed after the training
procedure, in which we minimize the velocity misfit $\mathcal{J}_v$. No additional parameters need to be estimated from data. For all test problems, we consider the same periodic square patch of ocean. 
\subsubsection{1D test problems}\label{subsubsec:1d_tests}
Recall that all training is based on steady-state DEM simulations with the triangular-shaped ocean velocities shown in figures~\ref{fig:velocity}. In a first generalization test, we instead use an ocean at rest, but a smooth wind profile that generates drag.  As shown in figure~\ref{fig:wind}, the steady-state DEM velocity data shows an excellent agreement with the continuum simulations across various concentrations.

\begin{figure*}[t]
    \centering
    \includegraphics[width=0.9\textwidth]{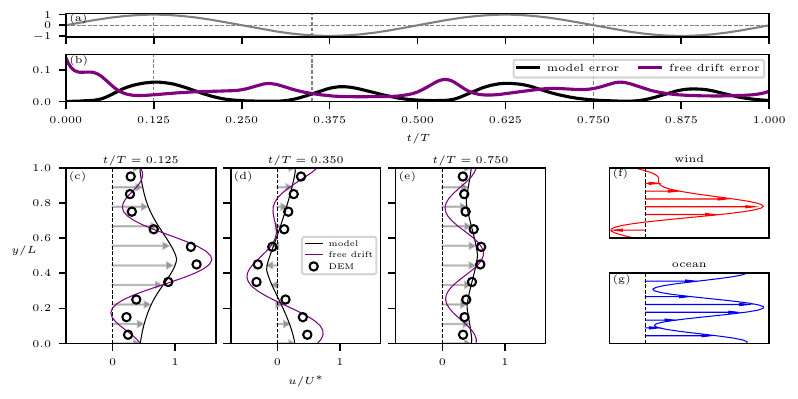}
    \caption{Unsteady test with concentration set to $A=0.875$. (a) Normalized amplitude of the wind velocity. (b) Difference between DEM and FEM model (black) and between DEM and free drift velocity (purple) (c-e) Comparison between velocity fields computed with our continuum model (black lines), the DEM (circles) and the free drift velocity (purple) for the one-dimensional unsteady test problem, at three different time instants. The velocities are non-dimensionalized with the equilibrium velocity $U^\ast = \sqrt{C_a\rho_a/(C_o\rho_o)}\,U_w$. (f) Shape of wind velocity profile when $U_w = 1$. (g) Shape of ocean velocity profile.}
    \label{fig:unsteady}
\end{figure*}

The next test challenges our model by considering unsteady wind currents and more complex spatial forcing profiles. For two concentrations, $A = 0.875$ and $A = 0.9$, we drive the floes with a horizontal ocean current $u_o(y)$ and a time-dependent horizontal wind field $u_w(y,t)$ over time $T = 1.4$ days. The profiles for $u_o$ and $u_w$ are chosen randomly for each value of $A$. We achieve this by writing the ocean and wind velocity profiles as linear combinations of two Fourier modes with random amplitudes and phase differences. These velocity profiles, despite being random, are set to enforce a maximum ocean velocity of $U_o = \SI{0.25}{\meter\per\second}$ and wind velocity of $U_w = \SI{20}{\meter\per\second}$. The ocean profile is kept constant in time, while the amplitude of the wind profile oscillates between $U_w$ and $-U_w$ over two periods.

In figure \ref{fig:unsteady}, we show the unsteady test results for $A = 0.875$, an ``unseen'' concentration that is not part of the training dataset. The random wind and ocean velocity profiles are shown in panels (f) and (g), respectively. In panels (c)-(e) we compare, at three instants in time, the DEM velocity with the horizontal velocity $u$ computed with our model. Additionally, we also present the free drift velocity $u_{\mathrm{FD}}$, which solves
\begin{align}
    \rho_oC_o|u_o - u_{\mathrm{FD}}|(u_o - u_{\mathrm{FD}}) + \rho_aC_a|u_w|u_w = 0.
\end{align}
The free drift velocity provides a reference to compare our model solution to: an accurate rheological model should, in principle, improve the approximation to the DEM provided by the free drift velocity, which ignores rheological and inertial effects. However, for low concentrations, achieving a better approximation with the model may be difficult because the rheological forces are small. To make the comparison more rigorous, in panel (b) we present the error difference with the DEM results for the model and free drift solutions. This error difference is defined as the average mean-squared difference and normalized with $U^\ast = \sqrt{C_a\rho_a/(C_o\rho_o)}\,U_w$. Additionally, panel (a) presents the amplitude of the wind velocity (this time-dependent amplitude multiplies the wind velocity profile in panel (f)). From these results we see that, despite the model coming close to the DEM data, at certain moments in time the free drift velocity provides a better approximation. In fact, we see that the instants in time where the free drift solution is more accurate correspond to points of amplitude 1 and -1. At these instants in time, the wind drag is large and dominates the sea ice dynamics, as seen in panel (c), where the DEM is very close to the free drift velocity.

\begin{figure*}[t]
    \centering
    \includegraphics[width=0.9\textwidth]{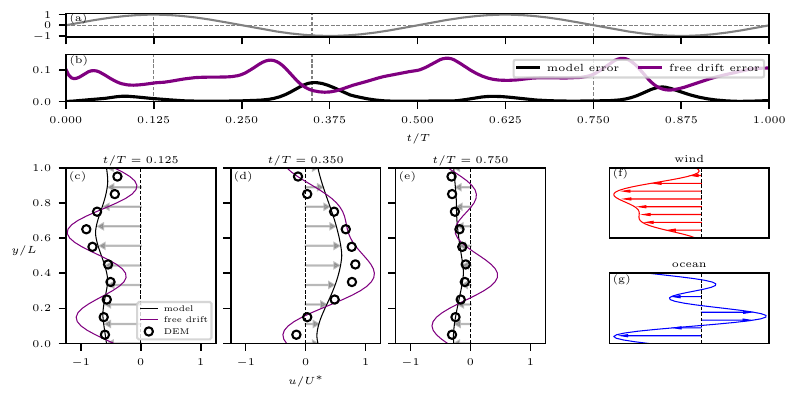}
    \caption{Unsteady test with concentration set to $A=0.9$. Same caption details as figure \ref{fig:unsteady}.}
    \label{fig:unsteady_2}
\end{figure*}

We repeat these computations for $A = 0.9$ with different wind and ocean velocity profiles. Unlike $A = 0.875$, for $A = 0.9$ the rheological forces are more dominant. Moreover, since $A = 0.9$ is a concentration included in the training data, we can expect to have a more accurate prediction of the DEM results. As seen in panel (b) of figure \ref{fig:unsteady_2}, this is the case: except for two short time intervals when the amplitude equals -1, the model error is very low. Therefore, figure \ref{fig:unsteady_2} indicates that our model is capable of reproducing the DEM's velocity field accurately in a setting largely differing from the training problems, especially for larger concentrations. This reinforces the claim that our framework for rheology inference has the capacity to discover the physics inherent to the system for sufficiently rich training datasets. 

\subsubsection{A 2D test problem}\label{subsubsec:2d_test}
\begin{figure}[t]
    \centering
    \includegraphics[width=0.35\linewidth]{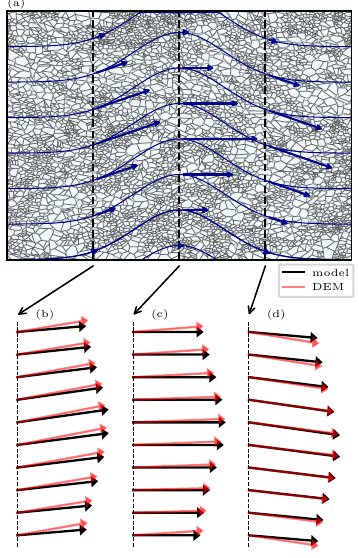}
    \caption{(a) Setup for the two dimensional test problem. The ocean velocity field follows the streamlines depicted in blue; its velocity vectors are depicted with arrows along three vertical cross sections. (b-d) Comparison of our model's (black) and the DEM's (red) velocity fields along three vertical cross sections at $x/L = 0.25, 0.5$ and $0.75$.}
    \label{fig:two_dim}
\end{figure}

The next test is two-dimensional. Sea ice is fundamentally compressible because ice floes may disperse or accumulate at different locations, changing the local concentration. The one-dimensional configuration we have used for training our model only allows a well-posed extension to two dimensions under the assumption of incompressibility. For high concentrations of sea ice that do not undergo any ridging or rafting, we expect an incompressible fluid model to be accurate. We test the validity of our model in two dimensions by solving an incompressible viscous fluid model whose shear viscosity is written in terms of $\psi_{\btheta}$, see Appendix \ref{appsubsec:incompressibility}. For a concentration of $A = 0.9$, we simulate the motion of ice floes with the DEM under an ocean velocity field that is no longer horizontal but follows the streamlines depicted in panel (a) of figure~\ref{fig:two_dim}. Along each vertical section, the velocity profile tangential to the streamlines is the triangle-shaped profile used for training the model (see (b) in figure~\ref{fig:setup}) with maximum velocity $U_o = \SI{0.5}{\meter\per\second}$. In panels (b)-(d) of figure~\ref{fig:unsteady}, we compare the velocity fields computed with the DEM and with the continuum model along $x/L = 0.25$, $0.5$ and $0.75$. Once again, our continuum model reproduces the DEM's velocity fields well. It preserves the symmetries in the mid-section $x = 0.5L$ of the ocean velocity field, such that the horizontal velocity $u$ is even and the vertical velocity $v$ is odd about $x = 0.5L$. Noticeably, the DEM does not preserve these symmetries, which may be due to the local sea ice concentration being slightly lower downstream of the domain than upstream. 
This redistribution of sea ice concentration can only be captured with a compressible continuum model.

\section{Discussion}

\subsection{Learned rheology} The numerical results in the previous section
demonstrate our framework's capacity to infer a shear rheology that can accurately reproduce the velocity fields computed with a complex DEM for sea ice. The resulting rheology provides valuable insight into the dynamics of sea ice: it reveals a transition from shear-thickening to shear-thinning behavior as the ice floe field becomes increasingly packed. This finding is in contrast to existing models for sea ice. Hibler's model, the state-of-the-art continuum model, predicts a plastic behavior for all concentrations \cite{hibler1979} as a consequence of ridging \cite{coon1974, rothrock1975}. Since, above a certain threshold, shear stress is independent of the strain rate in plastic materials, Hibler's model is an example of a shear-thinning rheology. Given that ridging is not considered in our DEM computations, we cannot expect our findings to be comparable to Hibler's model. Alternatively, collisional models for sea ice are built from a setup similar to ours. Early work derived a Newtonian (i.e.~linear) viscous model in which the viscosity increases, as in our case, with concentration \cite{shen1986, shen1987}. In these collisional models, internal stresses emerge from momentum transfer between ice floes via collisions. A more sophisticated derivation of the Newtonian viscous collisional model was recently provided \cite{toppaladoddi2025}. Collisional models, however, require a parameterization of forces due to floe-floe interactions which, in general, is very difficult to derive. A purely phenomenological rheology model that has been successful in modeling dense granular flows is the $\mu(I)$ model \cite{dacruz2005, pouliquen2006, jop2006}. Initial explorations found the $\mu(I)$ model capable of capturing DEM computations with disk-shaped ice floes \cite{herman2022}. The $\mu(I)$ has also demonstrated some accuracy in predicting DEM data from \emph{SubZero} \cite{dediego2024}. In fact, the results of \cite{dediego2024} translate into a shear-thickening viscosity for all concentrations. This shear-thickening viscosity, however, is incapable of capturing the sea ice dynamics observed in the DEM for very high concentrations. For $A = 0.95$, we find that the sea ice velocity profiles become increasingly curved as the ocean velocity increases, see panels (f)-(j) of figure \ref{fig:velocity}. Our results demonstrate that this trend can be predicted with a shear-thinning viscosity. Future work should explore the underlying cause of the transition from a shear-thickening to a shear-thinning viscosity.

\subsection{Learning from observations}
Rather than learning from DEM simulation data, one may want to use observation data of ice floe fields to learn the rheology. Such observations could come from satellites or ice floe trackers. Since obtaining stress data is extremely challenging \cite{ParnoPolashenksiParnoEtAl22}, a learning approach based on strain rates and stresses, i.e., using \eqref{eq:stress-misfit}, is generally infeasible. The velocity learning approach proposed in this work is a more suitable approach, as it does not require stress data. Instead, it minimizes the objective \eqref{eq:vel-misfit}, which compares ice velocity data with simulated ice velocities. The latter are the solution of the governing continuum equation, whose solution requires knowledge of the ocean and wind drag forces $\tau_o$ and $\tau_w$. Estimates of these forcing terms may be available from atmospheric and ocean measurements combined with modeling. An additional difficulty arises in the optimization of the NN weights; since stress data is unavailable, the two-step optimization procedure in which the stress misfit $\mathcal{J}_s$ is minimized to provide an initial guess for the velocity-based optimization cannot be carried out. We find that the velocity-based optimization converges with less accurate initial guesses, but often falls into local minima far from the global minimum if initialized with randomly-selected weights. For this reason, applications of this framework to observational data should consider the construction of good initial guesses or modifications of the optimization algorithm that make it more robust.

\subsection{Extensions}
In this work, we have chosen NNs for parameterization of the rheology function. Available software packages and the approximation properties of NNs, in particular in high dimensions, make this a natural choice. However, NNs are known to behave rather nonlinearly. This can make their training challenging, in particular when only moderate amounts of data are available and we are interested in well-converged minimizers.
An alternative to using NN parameterizations may be Gaussian process regression. In addition to the estimated function, Gaussian processes provide uncertainty estimates; that is, one could potentially say something about the regimes in which the data constrain the rheology function well and in which they do not. However, Gaussian processes tend to suffer more from the curse of dimensionality. This may become an issue when rheology functions aim at capturing more effects such as nonlocal behavior due to finite floe size, as discussed next.

Our framework for rheology inference can be extended to account for more complex physical phenomena. First, compressible effects, which are fundamental for accounting for spatial variations in concentration, can be introduced into the model by learning a bulk viscosity and an equation of state for the pressure. An extension to compressible models increases the amount of network parameters to be optimized and the size of the training dataset, requiring much larger computational resources. Second, our current approach assumes that the effective shear viscosity $\psi$ is local. Recent work suggests that nonlocal effects are important in the dynamics of granular media \cite{kamrin2012, berzi2024}. These effects could potentially be modeled by representing $\psi_{\btheta}$ with a convolutional neural network or an NN-based kernel operator. Third, our model assumes that the internal stresses in sea ice are, on average, isotropic. Anisotropy is known to emerge in the central ice pack as a consequence of the alignment of linear kinematic features \cite{wilchinsky2006, tsamados2013, dansereau2016}, although it is unclear whether such alignment occurs in our DEM computations. Future extensions of our framework could incorporate anisotropic effects by including a structure tensor that measures the degree of alignment of linear kinematic features. Finally, in our computations we deactivated the DEM's parameterizations for mechanical deformations of ice floes, such as fracturing, ridging, rafting, and welding. We excluded these phenomena because, given their complexity, the accuracy of their parameterizations is still to be verified. However, including mechanical deformation of ice floes in our computations is straightforward and can be explored in future work. Since mechanical deformations imply changes in floe size and thickness, the dependence of the rheology on these parameters would also have to be captured. Introducing additional parameters in the rheological law considerably increases the size of the training data set. This, in turn, makes the optimization more computationally demanding.

\section{Conclusions}

We have introduced a method for inferring an incompressible rheology for sea ice from data generated with a DEM that tracks individual ice floes. We represent the effective shear viscosity with an NN and incorporate the governing PDE into the training procedure. This allows us to train our NN using velocity data, as opposed to stress-strain-rate data, which are more noisy. We incorporate key physical and mathematical properties into the continuum model in the following ways. Firstly, by representing the rheology in terms of tensor invariants, we ensure that our model is isotropic and frame-indifferent. Secondly, by penalizing negative derivatives of the stress-strain map during the networks training, we guarantee the well-posedness of the model. Finally, by using custom-made activation functions in our NN, we guarantee that the effective shear viscosity is positive.

Our numerical results yield a highly nonlinear rheology for sea ice that transitions from a shear-thickening to a shear-thinning viscosity as the ice concentration increases. Additionally, we find a very strong dependence of the non-linear viscosity on the ice concentration.
We evaluate the generalizability of our model to problems that differ from the training setup. Our numerical results indicate high degrees of accuracy in reproducing the DEM's velocity fields on unseen ice concentrations, on unsteady problems driven by wind forces, and on two-dimensional configurations where compressible effects are small. These numerical tests demonstrate our approach's potential in discovering new continuum models for sea ice from data. We believe our data-driven approach represents a major step towards building new continuum sea ice models that can reproduce observed features with much higher accuracy than existing analytically-derived models.

\begin{acknowledgments}
The authors appreciate many helpful discussions with Dimitrios Giannakis and Mohammad Javad Latifi. They also would like to thank Skylar Gering for her support in using the Julia implementation of the sea ice DEM \emph{SubZero}.
Both authors were supported by the Multidisciplinary University Research Initiatives (MURI)
Program, Office of Naval Research (ONR) grant \#N00014-19-1-242. GS also appreciates support from the National Science Foundation under \#2343866 and \#2411229.
\end{acknowledgments}

\appendix

\section{Continuum model}\label{appsec:continuum_model}

\subsection{A general continuum model for sea ice}\label{appsubsec:general_continuum_model}

When representing sea ice as a two-dimensional continuum covering the ocean, conservation of momentum is expressed in terms of the sea ice velocity $\bu$ and the Cauchy stress tensor $\bsigma$ as a thickness-integrated balance of forces:
\begin{align}\label{eq:general_conserv_mom}
    \rho_i H \frac{\mathrm{D}\bu}{\mathrm{D}t} - \nabla\cdot \bsigma = \btau_o(\bu) + \btau_w.
\end{align}
Here, $\rho_i$ denotes the sea ice density, $H$ the ice thickness, and $\mathrm{D}/\mathrm{D}t$ the material derivative. The terms $\btau_o(\bu)$ and $\btau_w$ represent drag forces due to ocean and wind currents, respectively, and are given by
\begin{align*}
    \btau_o(\bu) &:= \rho_o C_o \|\bu_o - \bu\|(\bu_o-\bu),\\ 
    \btau_w(\bu) &:= \rho_a C_a \|\bu_w\|\bu_w,
\end{align*}
where $C_o$ and $C_a$ represents the drag coefficients, $\rho_o$ and $\rho_a$ the densities, and $\bu_o$ and $\bu_w$ the velocity fields of the ocean and wind, respectively. The scalar fields $\tau_o$ and $\tau_w$ defined in section \ref{sec:framework}, in \eqref{eq:defn_tau} and \eqref{eq:defn_tauw}, represent the horizontal components of $\btau_o$ and $\btau_w$, respectively, in the one-dimensional setting considered in section \ref{appsubsec:deriving_one_dim_model}. 

\subsection{Characterization of a local isotropic rheology} \label{appsubsec:isotropic_rheology}

As a starting point for restricting the functional form of a general rheology, we take an explicit and local dependence between the Cauchy stress tensor $\bsigma$ and the strain-rate tensor $\bD\bu$, given by
\begin{align*}
    \bsigma = -p\bI + \mathcal{C}(\bD\bu),
\end{align*}
and written in terms of a pressure $p$ and a function $\mathcal{C}:\RR^{2\times 2}\to \RR^{2\times 2}$. If we assume our material to be isotropic, that is, for any orthogonal matrix $\bO$, we have that,
\begin{align*}
    \mathcal{C}(\bO\bA\bO^\top) = \bO \mathcal{C}(\bA)\bO^\top \quad \forall \bA\in\RR^{2\times 2},
\end{align*}
then the function $\mathcal{C}$ can be characterized more precisely. One can show, using the Cayley-Hamilton theorem, that $\mathcal{C}$ is isotropic if and only if there exist two scalar functions $\psi_i:\RR^2\to\RR$ for $i = 1$ and 2 such that 
\begin{align*}
	\mathcal{C}\mathcal(\bA) = \psi_1(\iota_{\bA}) \bI + \psi_2(\iota_{\bA}) \bA,
\end{align*}
where $\iota_{\bA}\in\RR^2$ denotes the principal invariants of the matrix $\bA\in\RR^{2\times 2}$ \cite{spencer1958}. The functions $\psi_1$ and $\psi_2$ represent the effective bulk and shear viscosities, respectively. These can be written as 
\begin{align*}
	\iota_{\bA} := \left( \tr{\bA}, \| \bA \| \right),
\end{align*}
where $\tr{\bA}$ denotes the trace of $\bA$ and $\|\bA\|$ is given by 
\begin{align*}
    \|\bA\|^2 := \frac{1}{2}\tr{\left( \bA^2\right)}.
\end{align*}
Under the assumption of isotropy, discovering the rheology of sea ice is equivalent to finding the functions $\psi_1$ and $\psi_2$, together with an equation of state for $p$. The functions $\psi_1$ and $\psi_2$ will generally depend on other scalar fields relevant to sea ice, such as concentration $A$, thickness $H$, floe size, etc.

\subsection{Deriving the one-dimensional continuum model}\label{appsubsec:deriving_one_dim_model}

The one-dimensional problem we train our rheological model with is, when treated as a continuum, given by \eqref{eq:conservation_of_momentum}. This equation follows from \eqref{eq:general_conserv_mom} by setting the ocean and wind velocities to be purely horizontal and independent of the horizontal coordinate $x$, such that $\bu_o = (u_o,0)$ and $\bu_w = (u_w,0)$, with $u_o = u_o(y, t)$ and $u_w = u_w(y,t)$. If we further assume the sea ice concentration $A$ and $H$ to be spatially constant, as we do in this work, we can expect the sea ice velocity and pressure to also be horizontal and independent of $x$, such that $\bu = (u,0)$ with $u = u(y,t)$ and $p = p(y,t)$. Then, one can deduce \eqref{eq:conservation_of_momentum} from \eqref{eq:general_conserv_mom}, with $\tau_o = \rho_oC_o|u_o - u|(u_o - u)$, $\tau_o = \rho_aC_a|u_w|u_w$, and $\tau$, the shear stress, representing the off-diagonal component of the Cauchy stress tensor $\bsigma$. Under these conditions, $\|\bD\bu\|$ is the only nonzero principal invariant of $\bD\bu$ and is given by $|\srate|$, with the shear strain-rate $\srate$ representing the off-diagonal component $\bD\bu$. Moreover, $\psi_2$ is the only relevant rheological function in one dimension, such that the relationship between shear strain-rate and stress can be written as \eqref{eq:rheology}, with 
\begin{align}\label{eq:psi_definition}
    \psi(|\srate|, A) = \psi_2(0,|\srate|, A).
\end{align}
Above, we include the dependence of the rheology on the sea ice concentration $A$.

\subsection{Well-posedness of the steady one-dimensional model}\label{appsubsec:well_posedness}

The steady one-dimensional continuum model is given by \eqref{eq:conservation_of_momentum} without inertial terms; that is, by
\begin{align}\label{eq:steady_1D_conservation_of_momentum}
    - \partial_y\left(\psi\left(\frac{1}{2}\left|\partial_y u\right|, A\right) \partial_yu\right) = \rho_oC_o|u_o-u|(u_o - u) + \rho_aC_a|u_w|u_w.
\end{align}
Equation \eqref{eq:steady_1D_conservation_of_momentum} is solved repeatedly throughout the optimization of the neural network weights $\btheta$. To arrive at a robust optimization algorithm, it is crucial to restrict $\psi_{\btheta}$ to a class of functions that guarantees that the continuum model is well-posed. Therefore, it is important to understand under what conditions the continuum model is well-posed. If $\psi_{\btheta}$ is differentiable, as is our case, it is possible to show that a horizontal velocity $u$ solves \eqref{eq:steady_1D_conservation_of_momentum} if and only if it minimizes the energy functional
\begin{equation}\label{eq:energy_functional}
    \begin{split}
    \mathcal{F}(u) := \int_0^L\int_0^{\frac{1}{2}|\partial_yu|}2\psi(s, A)s\,\dd s\,\dd x
    +\frac{\rho_oC_o}{3} \int_0^L |u_o - u|^3\,\dd x - \rho_aC_a\int_0^L|u_w|u_wu\,\dd x
    \end{split}
\end{equation}
and this functional is convex.
The equivalence between \eqref{eq:steady_1D_conservation_of_momentum} and the minimization of $\mathcal{F}$ is established as follows. The directional (i.e.~G\^ateaux) derivative of $\mathcal{F}$ at $u$ in the direction of an arbitrary velocity profile $v$ is given by
\begin{equation}\label{eq:derivative_energy}
    \begin{split}
        \mathcal{F}'(u;v) = \int_0^L \psi\left(\frac{1}{2}|\partial_yu|,A\right) \partial_yu\,\partial_yv\,\dd x + \rho_oC_o\int_0^L |u_o - u|(u_o - u)v\,\dd x + \rho_aC_a\int_0^L |u_w|u_wv\,\dd x.
    \end{split}
\end{equation}
We note that \eqref{eq:steady_1D_conservation_of_momentum} can be rewritten in a variational form as $\mathcal{F}'(u;v) = 0$ for all $v$; indeed, by performing integration by parts, we find that $\mathcal{F}'(u;v) = 0$ can be rewritten as
\begin{equation}\label{eq:derivative_energy_2}
    \int_0^L \left( -\partial_y\left(\psi\left(\frac{1}{2}|\partial_yu|,A\right) \partial_yu\right) + \rho_oC_o |u_o - u|(u_o - u) + \rho_aC_a |u_w|u_w\right)v\,\dd x = 0.
\end{equation}
Now, if \eqref{eq:derivative_energy_2} holds for all $v$, we must have that \eqref{eq:steady_1D_conservation_of_momentum} holds by the fundamental lemma of the calculus of variations. Therefore, it suffices to show that $u$ minimizes $\mathcal{F}$ if and only if $\mathcal{F}'(u;v) = 0$ for all $v$. If $u$ is a minimizer of $\mathcal{F}$, it is clear that $\mathcal{F}'(u;v) = 0$ for all $v$. Conversely if we assume $\mathcal{F}'(u;v) = 0$ for all $v$, it follows that
\begin{equation}\label{eq:derivative_energy_3}
    \mathcal{F}'(u;v-u) = \lim_{t\to0}{\frac{\mathcal{F}\left(u + t(v-u)\right) - \mathcal{F}(u)}{t} = 0
}\end{equation}
for an arbitrary function $v$. In particular, if $t\in(0,1)$, we find that 
%
$    \mathcal{F}(u + t(v-u)) \leq t \mathcal{F}(v) + (1-t)\mathcal{F}(u)$
%
as a consequence of the convexity of $\mathcal{F}$. By setting this inequality into \eqref{eq:derivative_energy_3}, it is straightforward to show that $\mathcal{F}(u)\leq
\mathcal{F}(v)$ for all $v$, and the proof is complete.

If we additionally assume the energy functional $\mathcal{F}$ to be coercive, it can be shown that the minimizer must necessarily exist \cite{evans2010}. In practice, we find that enforcing convexity suffices to make the optimization robust. However, we ignore coercivity because it is unclear how to enforce it; the definition of coercivity depends on the function space on which the minimization problem is posed. The convexity of the rheological term in \eqref{eq:energy_functional} holds if and only if the map $s\mapsto \psi(s,A)s$ is monotonically increasing. Therefore, the monotonicity of $\psi(s,A)s$ is a sufficient condition to enforce the convexity of \eqref{eq:energy_functional}.

\subsection{Incompressible two-dimensional extension}\label{appsubsec:incompressibility}

If the continuum model is assumed to be compressible, it is not straightforward to extend our model to two dimensions. In the compressible case, the function $\mathcal{C}$ depends on $\psi_1$ and $\psi_2$, but our one-dimensional setup for training only reconstructs the dependence of $\psi_2$ on the second invariant of the strain-rate tensor $\bD\bu$. In addition, an equation of state for the pressure $p$ is required. However, if we assume the continuum model to be incompressible, such that
\begin{align}\label{eq:incompressibility}
	\nabla\cdot \bu = 0,
\end{align}
the only non-trivial invariant of the strain-rate tensor is the second invariant. Moreover, under the assumption of an incompressible medium, the isotropic component of the Cauchy stress tensor $\bsigma$ coincides with the Lagrange multiplier for the incompressibility constraint \eqref{eq:incompressibility}. This dispenses with the need to learn $\psi_1$ and an equation of state for $p$, leading to a well-defined extension of the continuum model to two dimensions. Under the assumption of incompressibility, the steady two-dimensional system is given by equation \eqref{eq:incompressibility} and
\begin{align*}
    \begin{split}
        - \nabla \cdot \left( 2 \psi(\|\bD\bu\|, A)\, \bD\bu\right) + \nabla p =\btau_o(\bu) + \btau_w.
    \end{split}
\end{align*}

\section{Optimization}\label{appsec:optim}

\subsection{Penalization to guarantee well-posedness}\label{appsubsec:penalty}

We can guarantee the convexity of the energy functional $\mathcal{F}(u)$, defined in \eqref{eq:energy_functional}, by enforcing that the function $s\mapsto \psi(s, A)s$ is monotonically increasing. In practice, we achieve this during the optimization of the network weights by penalizing negative values of the derivative of $\psi(s, A)s$. To this end, we define the penalty term
\begin{align*}
	\Pi(\btheta) := \int_{Q} \left| \min{\left\lbrace \frac{\partial}{\partial s}\left(\psi(|s|, A)\,s\right), 0 \right\rbrace}\right|^2\,\dd s\,\dd A.
\end{align*}
The function $\Pi(\btheta)$ penalizes points inside a set $Q\subset \RR_+\times[0,1]$ where the function $\psi(s, A)s$ is decreasing in $s$. 

\subsection{Optimization routine for training the neural network}\label{appsubsec:optimization_routine}

\begin{figure}[t]
\centering
\includegraphics[width=0.95\textwidth]{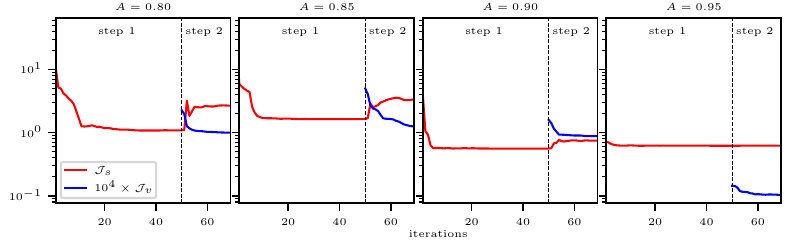}
\caption{Values of stress and the velocity misfit functions at different iterations of the optimization for each concentration $A$. The velocity misfit is multiplied by $10^4$ to scale its value up to the same order of magnitude as $\mathcal{J}_s$.}\label{fig:optimization}
\end{figure}

We compute the NN weight vector $\btheta$ that characterizes the effective shear viscosity $\psi_{\btheta}$ in two steps. First, we minimize the penalized stress misfit $\hat{\mathcal{J}}_s$ given by
\begin{align*}
    \hat{\mathcal{J}}_s(\btheta) :=  \mathcal{J}_s(\btheta) + \alpha_1\Pi(\btheta) + \alpha_2\|\btheta\|_{\ell^1}.
\end{align*}
Secondly, using the weight vector $\btheta$ computed in the previous step as an initial guess, we minimize the penalized velocity misfit
\begin{align*}
    \hat{\mathcal{J}}_v(\btheta) :=  \mathcal{J}_v(\btheta) + \beta_1\Pi(\btheta) + \beta_2\|\btheta\|_{\ell^1}.
\end{align*}

The training data consists of data points extracted from the DEM computations described in (ref. main text). In particular, we consider four sea ice concentrations $A = 0.8$, $0.85$, $0.9$, $0.95$. For each concentration, we find steady states with the DEM for seven maximum ocean velocities $U_o = 0.05$, $0.1$, $0.25$, $0.5$, $1$, $1.5$, $\SI{2}{\meter\per\second}$. For each steady state, we extract ten horizontal velocity data points and ten pairs of shear strain-rate/stress data points by averaging the DEM data over a grid and over the last $25\%$ of the time steps. 

We found it challenging to minimize the stress and velocity misfits over all four training concentrations at the same time. We therefore use the following approach, which resulted in a more reliable convergence and better fits. First, for each concentration, we minimize the velocity mismatch by following the two-step optimization routine. In this way, we compute four different effective shear viscosity models, one for each concentration. Then we compute the concentration-dependent effective viscosity $\psi_{\btheta} = \psi_{\btheta}(|\dot{\gamma}|, A)$ by fitting it to the four different viscosity models. This last step involves only points sampled from these four viscosity models, not DEM data. 

We minimize the penalized misfit functionals $\hat{\mathcal{J}}_s$ and $\hat{\mathcal{J}}_v$ with the LBFGS algorithm \cite{liu1989}. We run $20$ iterations to minimize $\hat{J}_s$ and $15$ for $\hat{J}_v$. The penalty parameters are set to $\alpha_1 = 10^{10}$, $\alpha_2 = 3\times 10^{-2}$, $\beta_1 = 10^{10}$ and $\beta_2 = 1.5\times 10^{-5}$. We find the algorithm to be insensitive to the monotonicity penalty parameters $\alpha_1$ and $\beta_1$, as long as these are sufficiently large to enforce the monotonicity. The optimization, however, is sensitive to the regularization penalty parameters  $\alpha_2$ and $\beta_2$. If these are two small, the resulting rheological models may exhibit rough features; conversely, if they are two large, the algorithm may converge to a linear model.

\begin{figure}[t]
\centering
\includegraphics[width=0.75 \textwidth]{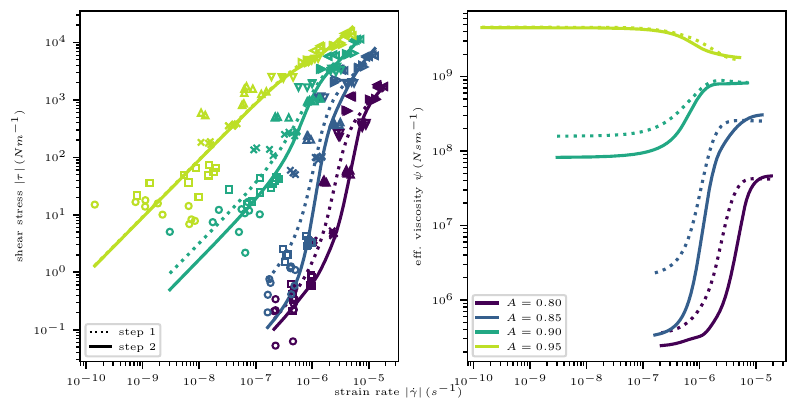}
\caption{(left) Shear strain-rate/stress relationship for the different concentrations (lines) learned from the DEM data (markers). (right) Effective viscosity $\psi_{\btheta}$ for different concentrations. We plot the rheological models computed from step 1 of the optimization algorithm (dotted line), where we minimize $\mathcal{J}_s$, and step 2 (straight line), where we minimize $\mathcal{J}_v$.}\label{fig:rheology}
\end{figure}

The values of the non-penalized functionals $\mathcal{J}_s$ and $\mathcal{J}_v$ over the optimization iterations are plotted in figure \ref{fig:optimization}. As expected, when optimizing for $\mathcal{J}_v$, a moderate increase in the value of $\mathcal{J}_s$ is observed. The rheological models learned in steps 1 and 2 of the optimization algorithm are plotted in figure~\ref{fig:rheology}.

\section{DEM implementation and parameters}\label{appsec:DEM}

\begin{table*}[t]
  \begin{center}
\def~{\hphantom{0}}
  \caption{Values for material parameters used in DEM computations and for the continuum model. Here, $C_o$ and $C_a$ are the drag coefficients for the ocean and wind, respectively, and $\rho_i$, $\rho_o$, and $\rho_a$ the ice, ocean water and air densities, respectively; these parameters are used in both the DEM and the continuum model. The Young's modulus $E$, Poisson's ratio $\nu$, and inter-floe friction coefficient $\mu^\ast$ are used exclusively in the calculation of collisional forces in the DEM, as described in \cite{manucharyan2022}.}
	\begin{tabular}{cccccccc}
		$C_o$ & $C_a$ & $\rho_i$ & $\rho_o$ & $\rho_a$ & $E$ & $\nu$ & $\mu^\ast$ \\
	\midrule
		$3\times 10^{-3}$ & $10^{-3}$ & \SI{900}{\kilo\gram\per\cubic\metre} & \SI{1027}{\kilo\gram\per\cubic\metre}& \SI{1.2}{\kilo\gram\per\cubic\metre} & \SI{6e6}{\pascal} & 0.3 & 0.2\\ \hline \\
	\end{tabular}	
  \label{tab:material_parameters}
  \end{center}
\end{table*}

The sea ice discrete element method we use to generate training data is \emph{SubZero} \cite{manucharyan2022}. While originally implemented in MATLAB \cite{MontemuroManucharyan23}, we use its re-implementation in the Julia language \cite{SubzeroJulia24} as it is faster and scales better in parallel. In table~\ref{tab:material_parameters}, we summarize important parameters used in the DEM (and partially also in the continuum model).
\emph{SubZero} generates floe fields with a Voronoi tessellation; after tessellating the domain, cells are removed at random, creating open patches of ocean, to achieve the desired sea ice concentration (see panel (c) in Figure 1). These Voronoi cells are based on a point cloud randomly drawn from a uniform distribution. Denoting by $y$ the areas of the resulting cells, it is known that in such a Voronoi tessellation, the number of cells $f_1(y)$ of size $y$ is well approximated by a generalized Gamma function with two parameters $a,b$:
\begin{equation}\label{eq:gamma-distribution}
f_1(y) = \frac{b^a}{\Gamma(a)} y^{a-1}\exp(-by).
\end{equation}
Typically used values in two dimensions are $a=3.61$ and $b=3.57$, \cite{ferenc2007size}. The distribution \eqref{eq:gamma-distribution} corresponds to a cloud of $p$ uniformly distributed points ($p\in\mathbb N$ is large) in a domain of area $p$, i.e., the expected cell size is $1$.

However, observational ice floe data show that the area of individual ice floes generally follows a power-law distribution \cite{buckley2024seasonal}, in stark contrast to \eqref{eq:gamma-distribution}. To be precise, the number of ice floes $f(y)$ with area $y$ is observed to satisfy
\begin{equation}\label{eq:f^tar}
    f(y) \approx Cy^{-m},
\end{equation}
where $C>0$ is a constant and the exponent $m>0$ is typically found to be between $1.75$ and $2$ for sea ice. To obtain a realistic floe size distribution (FSD) of cell areas from a Voronoi tesselation, we combine cells generated with denser point clouds across $K$ cell size distributions.
As detailed in \cite{Voronoi25}, we compute the fraction of the domain covered by points corresponding to the different cell size distributions by solving a non-negative nonlinear least squares problem for the domain fractions.
To obtain a reasonable mixing of differently sized Voronoi cells, we use tiles that each contain a different point density.
An example of the resulting ice floes, together with their floe size distribution, is shown in panel (b) of Figure~\ref{fig:setup}.

\bibliography{bibliography}

\end{document}